\documentclass[%
 aip,
 amsmath,amssymb,
 reprint,%
]{revtex4-1}

\usepackage{graphicx, xcolor}
\usepackage{dcolumn}
\usepackage{bm}
\usepackage{algorithmic}
\usepackage{algorithm}
\usepackage{multirow}
\usepackage[utf8]{inputenc}
\usepackage[T1]{fontenc}
\usepackage{mathptmx}
\usepackage{etoolbox}
\usepackage{subfig}
\makeatletter
\def\@email#1#2{%
 \endgroup
 \patchcmd{\titleblock@produce}
  {\frontmatter@RRAPformat}
  {\frontmatter@RRAPformat{\produce@RRAP{*#1\href{mailto:#2}{#2}}}\frontmatter@RRAPformat}
  {}{}
}%
\makeatother
\begin{document}

\preprint{AIP/123-QED}

\title[ ]{
Block Tensor Decomposition: A dual grid scheme with formal $O(N^3)$ scale for THC decomposition of molecular systems}

\author{Yueyang Zhang}%
\affiliation{The State Key Laboratory of Physical Chemistry of Solid Surfaces, Fujian Provincial Key Laboratory of Theoretical and Computational Chemistry, and College of Chemistry and Chemical Engineering, Xiamen University, Xiamen, Fujian 361005, China}

\author{Xuewei Xiong}
\affiliation{The State Key Laboratory of Physical Chemistry of Solid Surfaces, Fujian Provincial Key Laboratory of Theoretical and Computational Chemistry, and College of Chemistry and Chemical Engineering, Xiamen University, Xiamen, Fujian 361005, China}

\author{Wei Wu}
\affiliation{The State Key Laboratory of Physical Chemistry of Solid Surfaces, Fujian Provincial Key Laboratory of Theoretical and Computational Chemistry, and College of Chemistry and Chemical Engineering, Xiamen University, Xiamen, Fujian 361005, China}
 
\author{Peifeng Su*}
\email{supi@xmu.edu.cn}
\affiliation{The State Key Laboratory of Physical Chemistry of Solid Surfaces, Fujian Provincial Key Laboratory of Theoretical and Computational Chemistry, and College of Chemistry and Chemical Engineering, Xiamen University, Xiamen, Fujian 361005, China}

\date{\today}

\begin{abstract}
Accurate and fast treatment of electron-electron interactions remains a central challenge in electronic structure theory because post-Hartree-Fock  methods often suffered from the computational cost for 4-index electron repulsion integrals (ERIs). Low-rank approaches such as tensor hyper-contraction (THC) and interpolative separable density fitting (ISDF) have been proposed for Hartree-Fock exchange and correlation's calculations. Their application to molecular systems remains inefficient due to the construction of THC kernel whose time scale increases as quartic with the number of basis functions. In this work, we present an algorithm named block tensor decomposition (BTD) based on a dual grid scheme that combines Hilbert sort and pivoted Cholesky decomposition to generate compact interpolative grids, allowing strict $O(N^3)$ scaling for THC/ISDF kernel construction. The key parameters in BTD are optimized via differential evolution, balancing efficiency and accuracy. Furthermore, we apply BTD in scaled opposite-spin MP2 (SOS-MP2), leveraging sparse mapping in real space to achieve quadratic scaling for electron correlation calculation and linear scaling for exchange calculation. This work advances low-scaling THC/ISDF methodologies for molecular systems, offering a robust framework for efficient and accurate electronic structure computations.
\end{abstract}

\maketitle

\section{Introduction}
In electronic structure theory, the treatment of electron-electron interactions remains a fundamental challenge. Both Hartree-Fock (HF) and density functional theory (DFT) incorporate Coulomb and exchange interactions through a mean-field approximation. To account for correlation effects such as dispersion, post-HF/DFT methods including coupled cluster theory (CC)\cite{cc1,CC2,cc3} and many-body perturbation theory (MBPT)\cite{mbpt1,mbpt2,book} have been developed for more rigorous treatment of electron-electron interactions. However, these methods suffer from computational bottlenecks due to the 4-index electron repulsion integral (ERI) tensor:
\begin{equation}
(\mu\nu|\lambda\sigma) = \iint\frac{\chi_\mu(\bm{r}_1)\chi_\nu(\bm{r}_1)\chi_\lambda(\bm{r}_2)\chi_\sigma(\bm{r}_2)}{|\bm{r}_1 - \bm{r}_2|}\text{d}\bm{r}_1\text{d}\bm{r}_2,
\end{equation}
Here, $\mu,\nu,\lambda,\sigma$  are the indexes of orbitals or basis functions. resulting in a scaling of $O(N^5)$ or higher. To address this challenge, various low-rank techniques have been implemented, including Cholesky decomposition, \cite{cd1,cd2,cd2,cd4,Pedersen2009} resolution of identity (RI), \cite{Weigend2002,Hohenstein2010} chain of spheres for exchange (COSX), \cite{Neese_2009,Izsak2011,Izsak2013,Helmich-Paris2021} and tensor hyper-contraction (THC), \cite{Hohenstein_2012,thc1,thc2} interpolative separable density fitting (ISDF). \cite{isdf1,isdf2,isdf3,isdf4}

The low-rank approach can be classified into three kinds, including$:$ 1) Contraction of two third-order tensors or two third-order tensors with one matrix (RI and Cholesky decomposition);
\begin{equation}
    \begin{aligned}
           (\mu\nu|\lambda\sigma) \approx& \sum_M(\mu\nu|\tilde{M})(\tilde{M}|\lambda\sigma)\\
        =& \sum_{MN} (\mu\nu|M)(M|N)^{-1}(N|\lambda\sigma)
        \end{aligned}
    \label{RI-format}
\end{equation}
Here, $M,N$ are indexes of auxiliary function in RI or Cholesky decompsition. RI and Cholesky decomposition have been successfully used in self-consistent field calculations, especially, the calculation of the Coulomb term. \cite{Weigend2002} These methods also make great effort in the calculations of random phase approximation (RPA) and auxiliary field quantum Monte Carlo (AFQMC). \cite{RI-RPA,afqmc_r1,afqmc_r2}

\noindent 2) Contraction of two matrices and one third-order tensor (COSX and other pseudo-spectral methods); 
\begin{equation}
    (\mu\nu|\lambda\sigma) \approx \sum_{g}X_{\mu g}X_{\nu g}V^{g}_{\lambda \sigma} \approx \sum_{g}X_{\lambda g}X_{\sigma g}V^{g}_{\mu\nu}
\end{equation}
Here $g$ is index of real-space grid. The Hartree-Fock exchange time scale is reduced by COSX from $O(N^4)$ to $O(N^3)$ and introduces less deviation compared to RI whose time scale is still $O(N^4)$ for exchange calculation. \cite{Neese_2009,aCOSX}

\noindent 3) Contraction of five or six matrices (THC/ISDF).
\begin{equation}
    \begin{aligned}
            (\mu\nu|\lambda\sigma) \approx& \sum_{KL}X_{\mu K}X_{\nu K}X_{\lambda L}X_{\sigma L}V_{KL}\\
            =&\sum_{KL}X_{\mu K}X_{\nu K}X_{\lambda L}X_{\sigma L}\sum_{M}B_{KM}B_{LM}
        \end{aligned}
    \label{THC-format}
\end{equation}

\noindent Here, $K,L$ are indexes for THC's kernel, usually the grids in real space. The THC/ISDF approach, involving only second-order tensors, has been applied in post-SCF methods with time scale of $O(N^5)$ and larger one to reduce the time scale to $O(N^4)$.\cite{Hohenstein_2012, rrcc1, emo-thc-cc, pp-rpa} To decompose the 4c2e-integral in the format of Eq.\ref{THC-format}, many algorithms have been developed, such as least-square THC, \cite{thc1} randomized QR factorization (QRCP), \cite{isdf2} centroidal Voronoi tessellation (CVT), \cite{isdf4} machine learning-based method \cite{ml-isdf} and grid-free factorization. \cite{gfthc}

In periodic systems, THC/ISDF achieves $O(N^3)$ scaling and has been successfully combined with RPA and $GW$ methods, \cite{THC-RPA1,ISDFRPA,ml-isdf,isdf-rpa1} offering significant advantages over RI-based approaches that scale as $O(N^4)$. \cite{RI-RPA} However, for molecular systems, the generation of the $O(N^4)$ kernel renders THC inefficient for methods with lower computational complexity. Although recent advances have achieved $O(N^3)$ scaling for RPA, SOS-MP2, AFQMC and HF exchange calculations, \cite{isdf3,rpa_n3_1,rpa_n3_2,afqmc1,afqmc2,sos-cc2} this cubic scaling is conditional upon a key requirement, the number of non-negligible basis function pairs ($|\mu\nu\rangle$) is scale linearly with system size due to the locality of basis function (i.e., their spatial overlap decays rapidly with distance). Even when these conditions are met, the absolute computational cost remains proportional to the quartic order of the total number of basis functions. 

The motivation of this work is to develop an improved algorithm called block tensor decomposition (BTD) for the construction of THC kernel that achieves $O(N^3)$ scaling in the worst case while exhibiting cubic scaling with respect to the number of basis functions. It is noticed that Toddo has also developed a THC algorithm to reduce the SCF scale from $O(N^4)$ to formal $O(N^3)$ called the THC density difference SCF (dSCF). \cite{thc-dSCF} Unlike THC-dSCF, which makes THC approximation of 2c2e integral $(M|N)$, in this work a dual grid scheme is developed. Compared to previous work by Head-Gordon\cite{isdf3} in which K-means is used to generate interpolative grids and a shift is used to avoid the singularity of overlap-fitting matrix. our method utilizes Hilbert space filling curves for grid sorting combined with pivoted Cholesky decomposition\cite{pcd-THC} to eliminate redundant grid points. This approach yields more compact grid distributions and enables linear-scaling grid pruning through the combined use of Hilbert sorting and pivoted Cholesky decomposition. The efficiency and accuracy of BTD relied on some parameters. To achieve high efficiency and maintain acceptable accuracy, these parameters are optimized by differential evolution.  

To show the efficiency of BTD, we combined it with the scaled opposite spin second order Møller-Plesset perturbation theory (SOS-MP2)\cite{sos-mp2} formal $O(N^3)$ scaling. Furthermore, we use the sparse mapping to achieve a lower-scaling algorithm. Many low-scaling correlation methods have been developed as orbital-specific virtuals (OSV)\cite{osv1,osv2,osv3,osv4,osv5} and domain lone pair natural orbital (DLPNO). \cite{dlpno1,dlpno2} In Toddo's works, an atomic orbital-based SOS-MP2 is combined with THC to achieve time scale as $O(N^{2.6})$. \cite{thc-sos-mp21,thc-sos-mp22} In our work, the sparse map\cite{smap1,smap2,smap3} is done in real space by interpolative grids rather than atomic orbitals or local molecular orbitals. This sparsity can also be used in the generation of the BTD kernel to perform the most time-consuming step on a $O(N^2)$ scaling way.

The motivations and contents of this work are shown. 
\begin{enumerate}
    \item We propose a dual-grid THC/ISDF kernel generation method achieving strict $O(N^3)$ system-size scaling, effectively overcoming the computational bottleneck of conventional approaches.
    \item Pivoted Cholesky and Hilbert sorting are used to generate interpolative grids in an efficient way, which overcomes the unstable shortcoming of the pseudoinverse.
    \item Differential evolution is used to optimize BTD's parameters of grid and kernel generation for both high efficiency and acceptable accuracy.
    \item We integrate BTD with Hartree-Fock exchange and SOS-MP2 correlation through sparse mapping, enabling efficient electron correlation treatment with reduced computational cost.
\end{enumerate}
\section{Methodology}
\subsection{Introduction of THC-RI}
 In this subsection, we briefly introduce ISDF based on CVT for molecules such as THC-RI. In THC-RI, a set of grids $\{\bm{r}_K\}$ is used to generate THC kernel. These grids are generated based on weighted CVT or called K-means for each atom. The Becke weight is used and yields interpolation points that result in accurate compression for further calculation. More details of this algorithm could be find in Algorithm 1 in Head-Gordon's work. \cite{isdf3} The interpolative grids' position and weight are calculated as
\begin{equation}
    \begin{aligned}
        &\bm{r}_K = \frac{\sum_{g\in K}\bm{r}_gw_g}{\sum_{g\in K} w_g}\\
        &w_g = \sum_{g\in K} w_g
    \end{aligned}\label{clustering}
\end{equation}
$\bm{r}_g$ is grid belong to cluster whose center is $\bm{r}_K$ and $w_g$ is its weight. The half-kernel $B_{KM}$ is calculated by fitting 3c2e-integral $(\mu\nu|\tilde{M})$ in Eq.\ref{RI-format} as
\begin{equation}
    (\mu\nu|\tilde{M}) = \sum_{K}X_{\mu K}X_{\nu K}B_{KM}
\end{equation}
where $X_{\mu K} = \chi_\mu(\bm{r}_K)\sqrt{\omega_K}$ is the value of basis $\chi_\mu$ on grid $\bm{r}_K$ multiple the interpolative grid's weight. The fitting is done by least squares method as 
\begin{subequations}
    \begin{align}
        &B_{KM} = \sum_{L}S^{+}_{KL}A_{LM}\\
        &A_{LM} = \sum_{\mu\nu}X_{\mu L}X_{\nu L}(\mu\nu|\tilde{M})\label{most-time}\\
        &S^+_{KL} = [(\bm{S} +\delta\bm{I})^{-1}]_{KL}\\
        &S_{KL} = \sum_{\mu\nu}X_{\mu K}X_{\nu K}X_{\mu L}X_{\nu L}\label{overlap_S}
    \end{align}
\end{subequations}
The $\delta$ is a very small positive value less than $10^{-11}$ making $S_{KL}$ not singular. The most time-consuming step of THC-RI is Eq.\ref{most-time} with a time scale of $O(N^4)$. If shell-pair sparsity is used, the time scale of fitting calculation can be reduced to $O(N^3)$ for large systems. If other sparsity can be considered, the time consumption can be reduced further.

\subsection{Block tensor decomposition}
The $O(N^4)$ step makes the time consumption of THC-RI grow quartic with the basis function. To make a kernel generation scheme with a formal time scale $O(N^3)$, block tensor decomposition (BTD) is developed.  
\begin{figure*}
    \centering
    \includegraphics[width=1\linewidth]{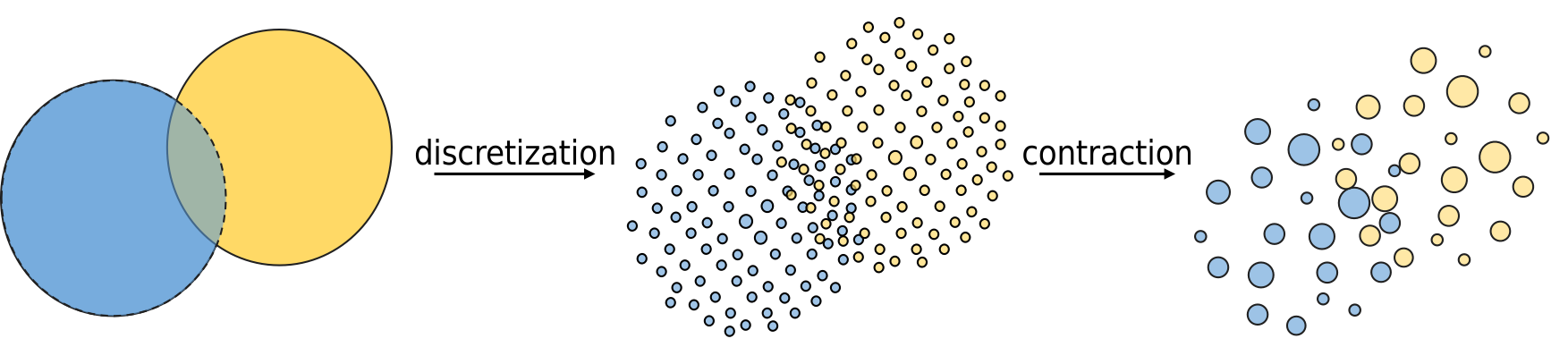}
    \caption{Conceptual diagram of BTD's approximation procedures.}
    \label{btd_cd}
\end{figure*}

The starting point of our method is that the continue electronic density $\rho(\bm{r})$ can be approximated by a set of discrete charges $\{Q_g\}$. The continuum-continuum Coulomb interaction can be approximated by the point-point interaction or the point-continuum interaction.
\begin{subequations}
\begin{align}
    &\rho(\bm{r}) \approx \sum_gQ_g\delta(\bm{r}-\bm{r}_g)\\
    &\iint\frac{\rho(\bm{r}_1)\rho(\bm{r}_2)}{|\bm{r}_1-\bm{r}_2|}\text{d}\bm{r}_1\text{d}\bm{r}_2\approx \sum_g \int\frac{\rho(\bm{r}_1)Q_g}{|\bm{r}_g-\bm{r}_1|}\text{d}\bm{r}_1\\
    &\iint\frac{\rho(\bm{r}_1)\rho(\bm{r}_2)}{|\bm{r}_1-\bm{r}_2|}\text{d}\bm{r}_1\text{d}\bm{r}_2\approx \sum_{gh}\frac{Q_gQ_h}{|\bm{r}_g-\bm{r}_h|}\label{dd-inter}
\end{align}
\end{subequations}

where $g,h$ are indexes of the point charges, $\delta(\bm{r})$ is the Dirac function and $\bm{r}_g$ is the coordination of charge $Q_g$. If point-point interaction in Eq.\ref{dd-inter} is used to approximate shell-pairs' interaction $(\mu\nu|\lambda\sigma)$ (as shown in Figure \ref{btd_cd}'s discretization step), the formula shown as below is gotten
\begin{subequations}
\begin{align}
    &(\mu\nu|\lambda\sigma) \approx\sum_{gh}Q^{\mu\nu}_gQ^{\lambda\sigma}_hV_{gh}\\
    &Q^{\mu\nu}_g = X_{\mu g}X_{\nu g}\\
    &V_{gh} = 1/|\bm{r}_g - \bm{r}_h|
\end{align}
\end{subequations}
These equations have a very similar format as THC shown in Eq.\ref{THC-format}. The difficulty in applying this approximation is that the number of grids used to discrete continue electronic density is always too huge. The number of grids can be reduced by finding more significant interpolative grids and contracting point charges in certain areas as one (as shown in Figure \ref{btd_cd}'s contraction step).   

In this work, the grids $\{\bm{r}_g\}$ used for the discretization are chosen as Lebedev grids. A set of sparse grids is generated from $\{\bm{r}_g\}$ as $\{\bm{r}_K\}$. Each $\bm{r}_K$ is the center of a cluster in $\{\bm{r}_g\}$. The details of $\{\bm{r}_K\}$'s generation are shown in section II C. The interactions in the sparse grids $\{\bm{r}_K\}$ and the interactions in the original grids $\{\bm{r}_g\}$ are connected by a contraction matrix $\chi_{Kg}$.
\begin{equation}
    V_{KL} = \sum_{gh}\xi_{Kg}\xi_{Lh}V_{gh}\\
\end{equation}
$\bm{\xi}$ can be seen as a contraction of the point charge on the original grids to the interpolative ones. This contraction shares the same idea of fast multipole method (FMM) in which a cluster of point charges or multipole interaction is approximated by the interaction of some multipoles on the center of this cluster. \cite{fmm1,fmm2,fmm3,bubble} In FMM, the contraction is done by multipole expansion and in our algorithm, ISDF type overlap fitting is used as belong.
\begin{subequations}
    \begin{align}
        &\xi_{Kg}=\sum_LS^+_{KL}\sum_{\mu\nu}Q_{g}^{\mu\nu}Q_{L}^{\mu\nu}\\
        &S^+_{KL} = [(\bm{S}+\delta\bm{I})^{-1}]_{KL}\\
        &S_{KL}=\sum_{\mu\nu}Q^{\mu\nu}_KQ^{\mu\nu}_L\label{SKL}
    \end{align}
\end{subequations}

The algorithm used to generate the THC kernel has a time scale of $O(N_g^2\times N_K)$ where $N_g$ is the number of original grids and $N_K$ is the number of sparse grids. Due to the huge value of $N_g$, the time consumption is still high. To solve this problem, RI is used to decompose the point-point interaction as the contraction of two point-continuum interactions.
\begin{subequations}
    \begin{align}
        &V_{gh} = 1/|\bm{r}_g - \bm{r}_h| \approx \sum_M B_{gM}B_{hM}\label{btd_ri}\\
        &B_{gM} = \sum_{N}(M|N)^{-1/2}\int\phi_N(\bm{r})/|\bm{r} - \bm{r}_g|\text{d}\bm{r}
    \end{align}
\end{subequations}
where $\phi_N$ is RI auxiliary function. In this case, the time scale is reduced to $O(N_g\times N_K\times N_\text{aux})$ where $N_\text{aux}$ is the number of auxiliary functions. The calculation of $\xi_{Kg}$ whose time scale is $O(N_g\times N_K^2)$ and by using RI the time scale is reduced to $O(N_g\times N_K\times N_\text{aux})$ and the calculation of the contracted half-kernel $B_{KM}$ is carried out as follows.
\begin{subequations}
    \begin{align}
        &B_{KM} = \sum_{L}S_{KL}^+A_{LM}\label{HK_1}\\
        &A_{LM} = \sum_{g}S_{gL}B_{gM}\label{HK_2}\\
        &S_{gL} = (\sum_\mu X_{\mu g}X_{\mu L})^2\label{HK_3}
    \end{align}
\end{subequations}
If the real space sparsity of the basis functions is used, the time consumption is reduced to about $O(N_g\times N_\text{aux})$ for the calculation of $A_{LM}$ and $S_{gL}$. The sparse map between $\bm{r}_g$ and $\bm{r}_K$ is determined as 
\begin{equation*}
    L(\bm{r}_g\rightarrow\bm{r}_K) = L(\bm{r}_g\rightarrow\chi_\mu)\cup L(\chi_\mu\rightarrow\bm{r}_K)
\end{equation*}
The notation follows the Neese's works. \cite{smap1} To perform the sparse calculation efficiently, $\{\bm{r}_g\}$ and $\{\bm{r}_K\}$ are divided into some blocks of certain size and the sparse map is made by blocks instead of by using each grid.

In the above discussions, an algorithm based on dual grids is developed for the decomposition of the THC. By combining BTD with the sparse map and RI, the time consumption can be further reduced. The problem remained is how to generate sparse interpolative grids from dense ones.
\subsection{Grids generation scheme based on Hilbert curve and pivoted Cholesky decomposition}
\begin{figure}[h]
    \centering
    \includegraphics[width=1\linewidth]{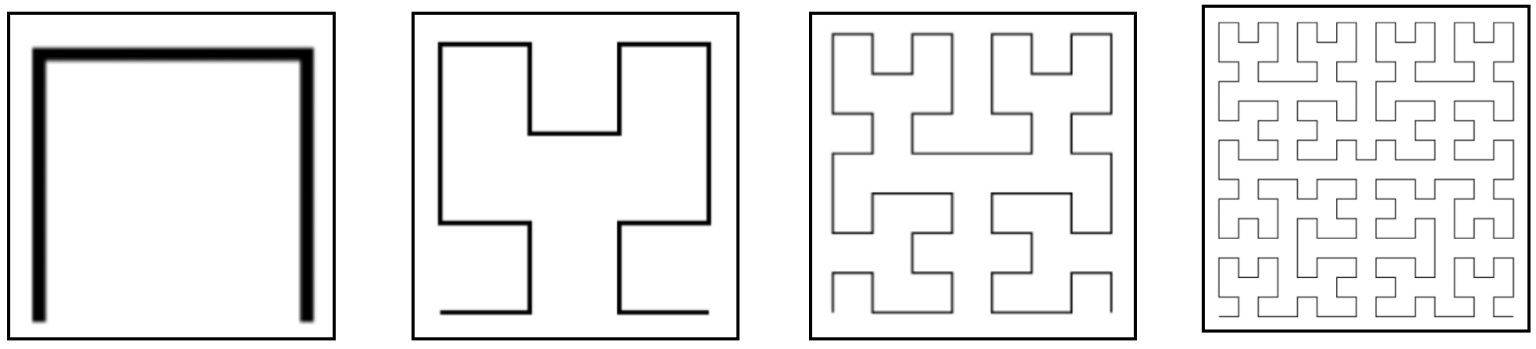}
    \caption{2D Hilbert curve in order from 1 to 4.}
    \label{hilbert}
\end{figure}

In this work, we generate sparse grids for overlap fitting by using Hilbert sorting, a method that arranges all 3D grids along a one-dimensional curve known as the Hilbert curve. The Hilbert curve preserves spatial proximity, meaning that points close to each other in 3D space remain close when projected onto the 1D curve. A 2D example of the Hilbert curve is illustrated in Figure \ref{hilbert}. In particular, Hilbert sorting has been successfully applied in semi-numerical algorithms for local hybrid density functional calculations. \cite{hilbert-cosx}

The Hilbert curve is cut into $N_K$ blocks as $\{\bm{r}_{K\times n_\text{block}}:\bm{r}_{\text{min}[N_g,(K+1)\times n_\text{block}]}\}$ where $K$ is the index of block from 1 to $N_K$ and $n_\text{block}$ is the size of block. The coordination and weight of the sparse grid are determined by Eq.\ref{clustering}.Compared to K-means clustering, the Hilbert curve approach offers a distinct advantage. While the K-means results depend on initial configurations and often converge to local minima, Hilbert sorting guarantees deterministic partitioning, and its results have a multilayer structure, which is useful for further pruning in this work.

The sparse grids generated from Hilbert sorting are still redundant and the overlap fitting matrix $S_{KL}$ is singular. To eliminate redundant grids, the pivoted Cholesky decomposition is used. This method selects the column with the largest diagonal element for decomposition and terminates when the maximum diagonal element falls below a predefined threshold $\epsilon$. Based on the property of the Hilbert curve, a linear-scaling pruning scheme is developed compared to the scale $O(N^3)$ of the Cholesky decomposition. In this scheme, the grids are divided into blocks of a certain size, and the pivoted Cholesky decomposition is done in each block to remove redundant grids. After this pruning, the remaining grids will be divided into blocks and pruned again. The spatial preservation of the Hilbert curve ensures that pruned grids maintain their proximity relationships. In this way, an iterative algorithm is designed as shown in Figure \ref{pCD}. The size of the block $n_\text{block}$ is related to the time consumption and the number of grids after pruning. In our test, the size of 1024 is efficient. 

In the BTD calculation, some parameters shown in Table \ref{para} are related to efficiency and accuracy. To make the calculation efficient and maintain accuracy, a parameter adaptive differential evolution algorithm named JADE is used to train these parameters. \cite{jade} The results are shown in section III B.
\begin{table}[H]
    \centering
    \caption{Parameters of BTD to be trained.}
    \begin{tabular}{|c|c|l|}
    \hline
    ID&symbols&$\qquad\qquad\qquad$variables\\
    \hline
    1&$\{l_1,l_2,l_3,l_4,l_5\}$&the number of angular grids for dense ones\\
    \hline
    2&$\{d_1,d_2,d_3,d_4,d_5\}$&the number of radial grids for dense ones\\
    \hline
    3&factor&the number of sparse grids is factor$\times N_\text{aux}$\\
    \hline
    4&$\epsilon$&the threshold of pivoted Cholesky prnuing\\
    \hline
    \end{tabular}\label{para}
\end{table}

\begin{figure}[h]
    \flushleft
    \begin{equation*}
        \begin{aligned}
            &\textbf{REQUIRED: }n_\text{block},\{\bm{r}_g\},\epsilon\\
            &\textbf{ENSURE: }\{\bm{r}_g\}\text{ after pruning}\\
            &\text{sort the parent grids }\{\bm{r}_g\}\text{ by Hilbert curve}\\
            &\text{labeled }\{\bm{r}_g\}\text{ as }\{\bm{r}_g^{(0)}\}\\
            &\textbf{DO}\\
            &\quad\text{divide grids into block as }\{\bm{r}_{m\times n_\text{block}}:\bm{r}_{(m+1)\times n_\text{block}}\}\\
            &\quad \textbf{FOR }m\text{ in blocks}\\
            &\qquad \text{calculate }S_{gh} = \sum_{\mu\nu}\omega_g\omega_hC_{\mu\nu}^gC_{\mu\nu}^h\\
            &\qquad \text{do }\textbf{dpstrf}\text{ with cutoff }\epsilon\\
            &\qquad\text{select grids from Cholesky as }\{\bm{r}^{(i+1)}\}\\
            &\quad\textbf{END FOR}\\
            &\textbf{WHILE}(\text{num}(\bm{r}^{(i)})\neq\text{num}(\bm{r}^{(i+1)}))
        \end{aligned}
    \end{equation*}
    \caption{The algorithm of pruning redundant grids by diagonal-blocked pivoted Cholesky decomposition in an iterative way.}
    \label{pCD}
\end{figure}

In summary, the time scale of BTD is $O(N^2-N^3)$ compared to that of THC-RI as $O(N^3-N^4)$. The algorithm is divided into two parts, namely, generation of sparse grids $\{\bm{r}_K\}$ and calculation of the THC kernel $B_{KM}$. The time scale for grid generation is between $O(N)$ and $O(N\text{log}(N))$ (time scale of Hilbert sorting) by the discussion in section II C. The algorithm for calculating the THC kernel and the time scale for each step are summarized in Figure \ref{BTD_algorithm}. 
\begin{figure}
    \centering
    \begin{equation*}
        \begin{aligned}
            &\textbf{REQUIRED: }\{\bm{r}_g\},\ \{\bm{r}_K\}, \{\phi_M(\bm{r})\}\\
            &\textbf{ENSURE: }B_{KM},\ V_{KL}\\
            &\text{calculate }S_{KL}\text{ by Eq.\ref{overlap_S}}&&O(N_K^2N_\text{bas})\rightarrow O(N_K)\\
            &\text{calculate inversion of }S_{KL}\text{ as }S_{KL}^+&&O(N_K^3)\\
            &\text{calculate }(M|N)^{-1/2}\text{ by Cholesky}&&O(N_\text{aux}^3)\\
            &\textbf{FOR }\bm{r}_g\text{ in dense grids}\\
            &\quad\text{calculate }X_{\mu g}\text{ for }L(\chi_\mu\rightarrow\bm{r}_g)&&O(N_gN_\text{bas})\rightarrow O(N_g)\\
            &\quad\text{calculate }A_{gM}\text{ as 2c1e-integral}&&O(N_gN_\text{aux})\\
            &\quad\textbf{FOR }\bm{r}_K\text{ in }L(\bm{r}_K\rightarrow\bm{r}_g)\\
            &\qquad \text{calculate }X_{\mu K}\text{ for }L(\chi_\mu\rightarrow\bm{r}_K)&&O(N_KN_\text{bas})\rightarrow O(N_K)\\
            &\qquad \text{calculate }S_{gL}\ \text{by Eq. \ref{HK_2}}&&O(N_gN_K N_\text{bas})\rightarrow O(N_g)\\
            &\qquad\text{calculate }A_{LM}\ \text{by Eq. \ref{HK_3}}&&O(N_gN_KN_\text{aux})\rightarrow O(N_KN_\text{aux})\\
            &\quad\textbf{END FOR}\\
            &\textbf{END FOR}\\
            &\text{calculate }B_{KM}\ \text{by Eq. \ref{HK_1}}&&O(N^2_KN_\text{aux})\\
            &\text{calculate }B_{KN}(M|N)^{-1/2}\rightarrow B_{KM}&&O(N_KN^2_\text{aux})\\
            &\text{calculate }V_{KL}\text{ by Eq.\ref{btd_ri}}&&O(N^2_KN_\text{aux})
        \end{aligned}
    \end{equation*}
    \caption{The algorithm of calculation for THC kernel. The time scale of each step is shown on the right hand and the asymptotic time scale is shown by right arrow.}
    \label{BTD_algorithm}
\end{figure}
\subsection{Implementation of exchange and SOS-MP2}

In this work, BTD-based THC is combined with Hartree-Fock exchange and SOS-MP2 correlation, to reduce the time scales to formal $O(N^3)$. To make the calculation efficient and approach linear scaling, the sparse grids are divided into certain blocks by Hilbert sorting or octree. \cite{octree1,octree2} prescreening is done to remove pairs of no-interaction blocks. Using prescreening, the time scale of Hartree-Fock exchange and SOS-MP2 correlation can be reduced to $O(N)$ and $O(N^2)$, respectively.

The exchange matrix is calculated by Eq.\ref{exchange} on a time scale of $O(N_K^2N_\text{bas})$.
\begin{subequations}
    \begin{align}
        &K_{\mu\nu} = \sum_{KL}X_{\mu K}X_{\nu L}U_{KL}\\
        &U_{KL} = V_{KL}\sum_{\mu\nu}X_{\mu K}X_{\nu L}P_{\mu\nu}
    \end{align}\label{exchange}
\end{subequations}
where $P_{\mu\nu}$ is matrix element of density matrix. The prescreening is performed by estimating $X_{\mu K}X_{\nu L}P_{\mu\nu}$ for the block pair of $A =\{\bm{r}_K\}$ and $B=\{\bm{r}_L\}$ as follows.
\begin{equation}
    \begin{aligned}
        &\text{THR}_1\le P_\text{max}\times X_{A,\text{max}}\times X_{B,\text{max}}\\
        &P_\text{max} = \text{max}\{|P_{\mu\nu}|;\ \mu\in L(\chi_\mu\rightarrow A),\ \nu\in L(\chi_\nu\rightarrow B) \}\\
        &X_{\mu,\max} = \max\{|X_{\mu K}|; \bm{r}_K\in A\}\\
        &X_{\nu,\max} = \max\{|X_{\nu L}|; \bm{r}_L\in B\}
    \end{aligned}
\end{equation}
A screening for each basis function can be performed to further reduce the number of interaction pairs.
\begin{subequations}
    \begin{align}
        &\text{THR}_2\le\max(T_A,\ T_B)\\
        &T_A = \max\{X_{\mu,\max}P_{\mu,\max}X_{B,\max};\ \mu\in L(\chi_\mu\rightarrow A)\}\\
        &T_B = \max\{X_{A,\max}P_{\nu,\max}X_{\nu,\max};\ \nu\in L(\chi_\nu\rightarrow B)\}\\
        &P_{\mu,\text{max}} = \text{max}\{|P_{\mu\nu}|;\ \nu\in L(\chi_\nu\rightarrow B) \}\\
        &P_{\nu,\text{max}} = \text{max}\{|P_{\mu\nu}|;\ \mu\in L(\chi_\mu\rightarrow A) \}\\
        &X_{\mu,\max} = \max\{|X_{\mu K}|; \bm{r}_K\in A\}\\
        &X_{\nu,\max} = \max\{|X_{\nu L}|; \bm{r}_L\in B\}
    \end{align}
\end{subequations}
Using the local property of electronic density, the density matrix element $P_{\mu\nu}$ decays exponentially as the basis function's behavior and makes the screening efficient enough to make the calculation linear scaling. In the implementation in this work, the selection is done by copying the density matrix from global to local, creating a bottleneck with the $O(N^2)$ time scale.

SOS-MP2 is an approximation of MP2 by just considering the opposite-spin part. The SOS-MP2 correlation energy is shown in Eq.\ref{SOS-MP2}.
\begin{equation}
    E_\text{SOS-MP2}^\text{corr} = \text{f}\times\sum_{ijab} \frac{(ia|jb)(ia|jb)}{\epsilon_i+\epsilon_j -\epsilon_a-\epsilon_b}\label{SOS-MP2}
\end{equation}
where f is the scale coefficient , $i,j$ are indexes of occupied orbitals, $a,b$ are indexes of virtual orbitals, $(ia|jb)$ is ERI in the molecular orbital representation and $\epsilon$ is the orbital energy. Using the Laplace transformation\cite{ls-mp21,ls-mp22,ls-mp23} and RI, the SOS-MP2 correlation energy can be reduced to $O(N^4)$. In our scheme, the calculation is done on the time scale of $O(N^3)$.
\begin{subequations}
    \begin{align}
        &E_\text{SOS-MP2}^\text{corr} = -\text{f}\times\sum_{n}w_n\sum_{MN}\Pi_{MN}(\tau_n)\Pi_{MN}(\tau_n)\\
        &\Pi_{MN}(\tau) = \sum_{KL}B_{KM}B_{LN}P_{KL}(\tau)\\
        &P_{KL}(\tau) = -G^<_{KL}(\tau)G^{>}_{KL}(\tau)\\
         &G^{>}_{KL}(\tau) = \sum_{a}\exp(-\tau\epsilon_a)X_{aK}X_{aL}\\
        &G^<_{KL}(\tau) = -\sum_{i}\exp(\tau\epsilon_i)X_{iK}X_{iL}\\
        &X_{iK} = \sum_{\mu}X_{\mu K}C_{i\mu}\\
        &X_{aK} = \sum_{\mu}X_{\mu K}C_{a\mu}
    \end{align}
\end{subequations}
Here $K'$ is the index of THC grid, $\tau_n$ is the Laplace sampling point,$w_n$ is the weight of the Laplace transformation, and $C_{i\mu}$, $C_{a\mu}$ are the molecular orbital coefficients for occupied space and virtual space, respectively. The time scale is $O(N^2_KN_\text{aux})$.

The time consumption of SOS-MP2 can be reduced to $\min\{O(N^2_\text{aux}N_K),O(N_K^2)\}$ by prescreening the polarization propagator $P_{KL}(\tau)$. The Green's functions in the basis function representation are defined as follows for the occupied and virtual space, respectively.
\begin{subequations}
    \begin{align}
        &G_{\mu\nu}^>(\tau) =  \sum_{a}C_{a\mu}C_{a\nu}\exp(-\tau\epsilon_a)\\
        &G_{\mu\nu}^<(\tau) =  -\sum_{i}C_{i\mu}C_{i\nu}\exp(\tau\epsilon_i)
    \end{align}
\end{subequations}
In the limitation $\tau\rightarrow0$, $-G_{\mu\nu}^<(0) = P_{\mu\nu}$ is just the density matrix. The prescreening is performed in a similar way as exchange by estimating $X_{\mu K}\times X_{\nu L}\times\max(|G_{\mu\nu}^>|,|G_{\mu\nu}^<|)$. Except for the calculation of $G_{\mu\nu}^{>,<}$ whose time consumption can be ignored, the calculation of $P_{\mu\nu}(\tau)$ is linear scaling, and $P_{\mu\nu}(\tau)$ is sparse. The time scale can be reduced to $O(N_K^2)$.
\begin{figure}[H]
    \centering
    \includegraphics[width=1\linewidth]{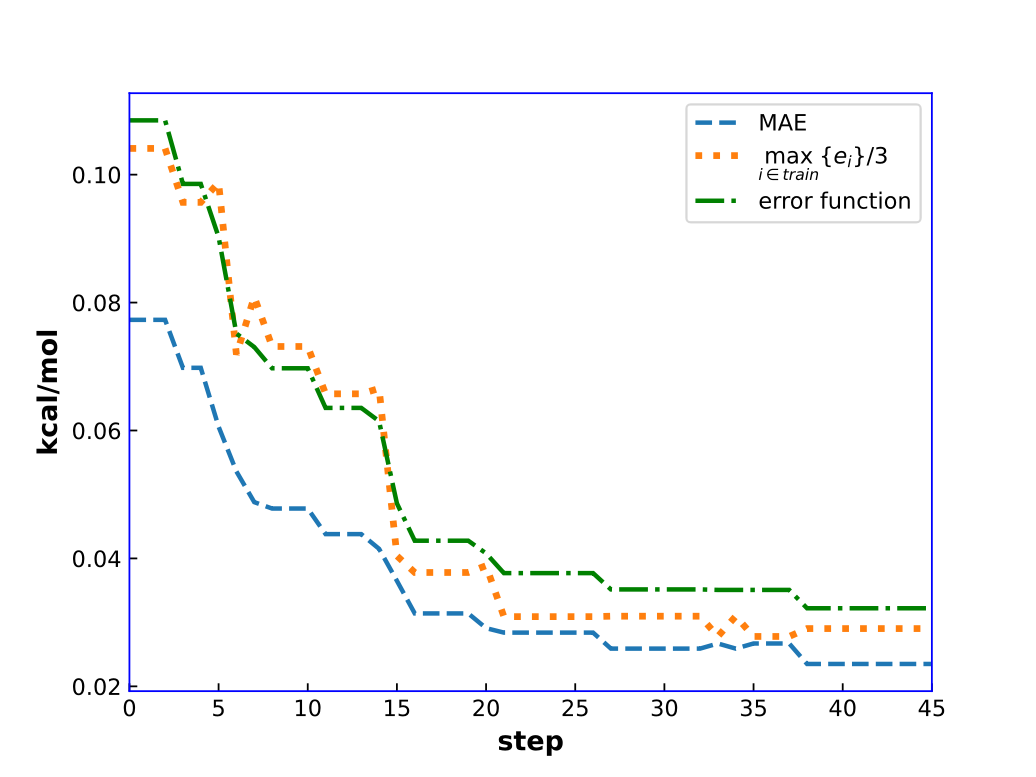}
    \caption{Training process with 20 samples one cycle. Calculation begins with random guesses and runs 45 steps.}
    \label{fig:enter-label}
\end{figure}
\begin{figure*}
    \centering
     \subfloat[cluster of waters]{
        \includegraphics[scale=0.35]{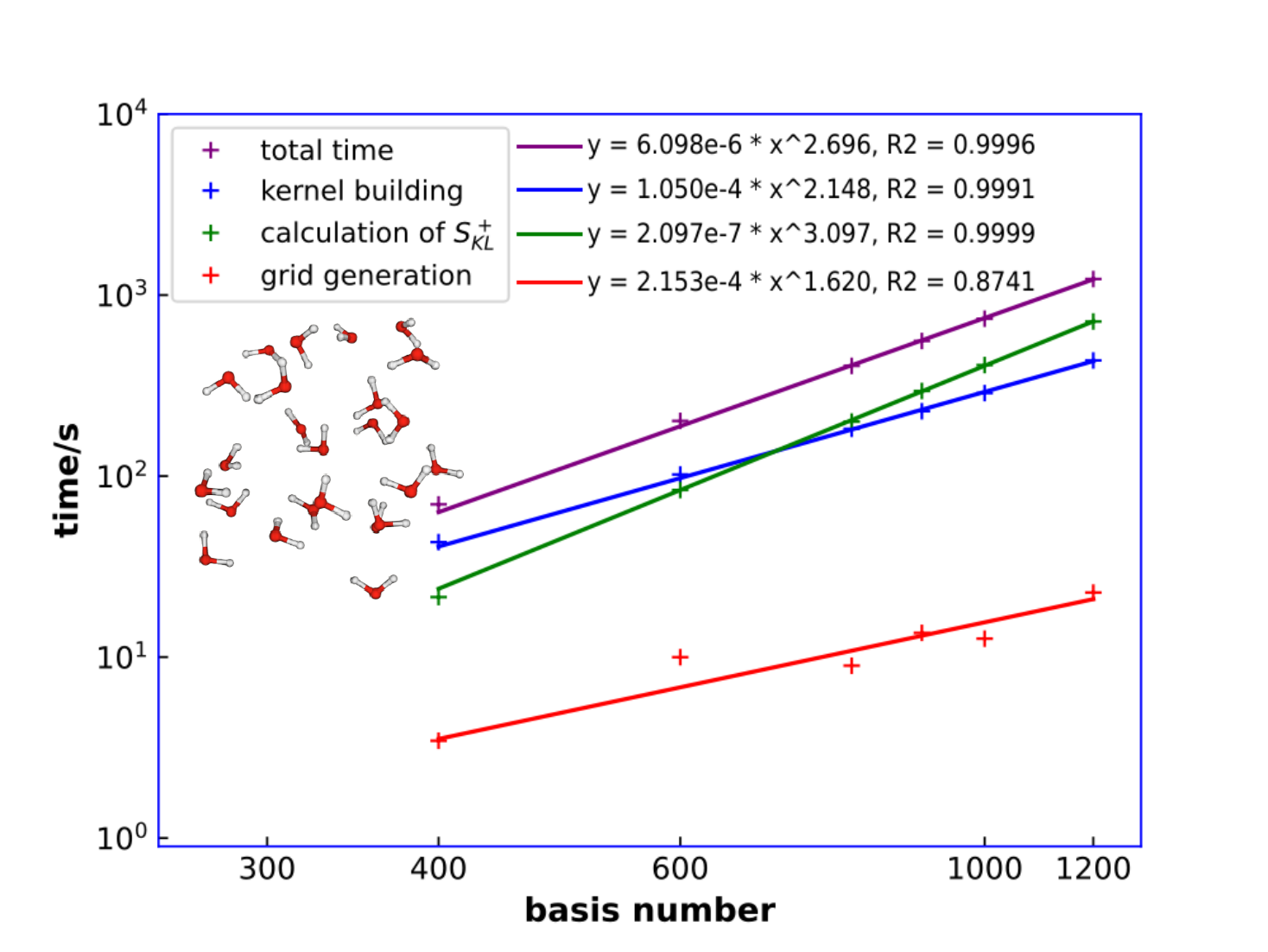}
      }
      \subfloat[chain of glycines]{
        \includegraphics[scale=0.35]{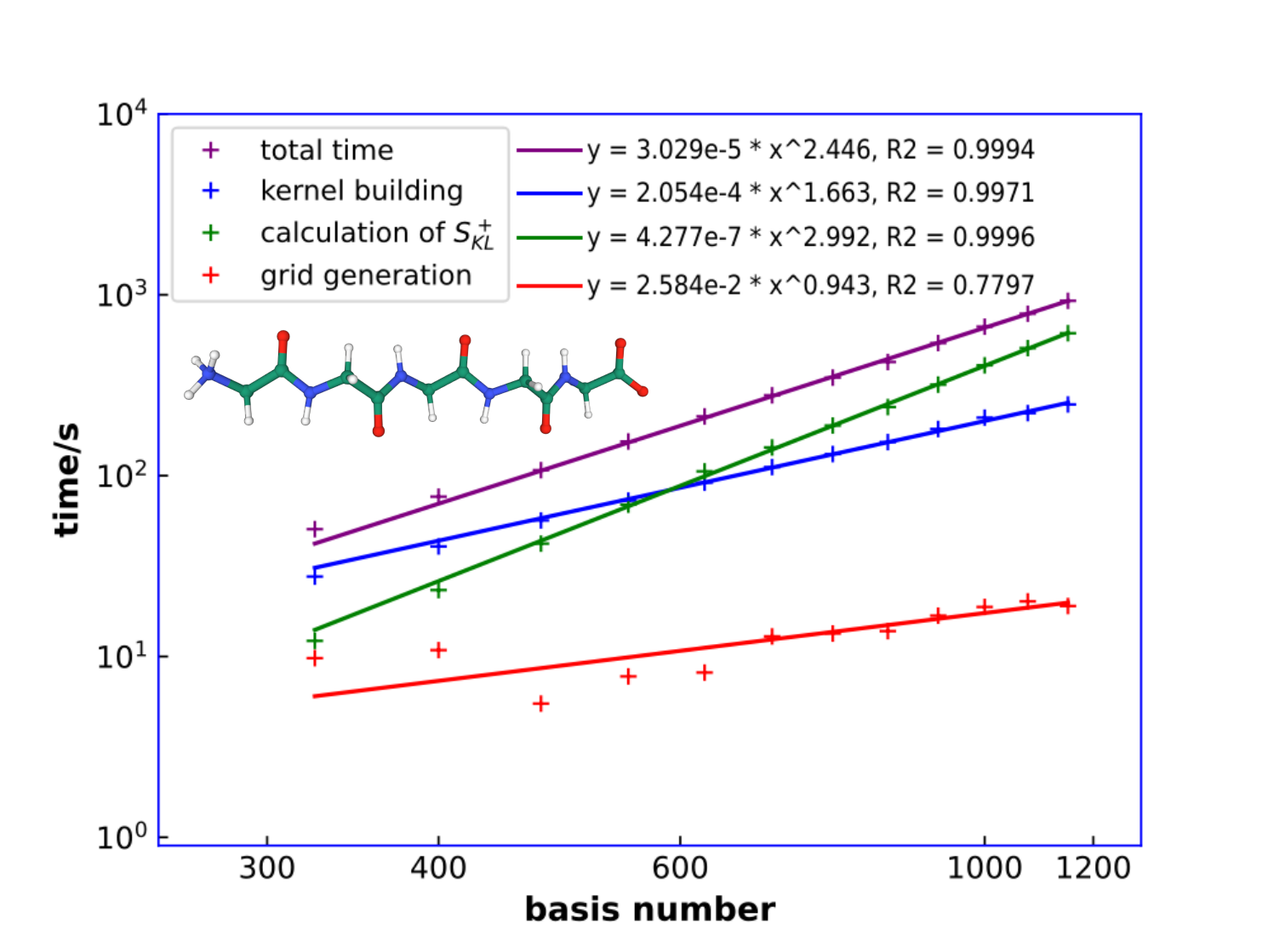}
      }
      \caption{The scaling of individual steps in BTD calculations on water clusters and glycines chains. All calculations use the cc-pVDZ basis set and the cc-pVDZ-RI auxiliary basis set.}
      \label{TIME_SCALE}
\end{figure*}
 \section{Results and discussions}
 The method of block tensor decomposition is implemented in the XEDA@XACS quantum chemistry program. Electron integral calculation is supported by Libxint. All calculations were done on Intel(R) Xeon(R) Gold 6226R CPU @ 2.90GHz. The cc-PVDZ is used for the calculations. For exchange calculation, the JKfit auxiliary functions cc-PVDZ-JKfit are used and for correlation's calculation, cc-PVDZ-RI is used.

\subsection{BTD kernel generation}
 To show the time scale of BTD's kernel generation and better understand the computational bottleneck, calculations were done for 1D and 3D systems as glycine chains and water clusters, respectively. The tests were done under cc-PVDZ and cc-PVDZ-JKfit is used as an auxiliary function with one CPU core. The numbers of waters in water clusters are 6, 24, 32, 36, 40, 48. The numbers of monomers in the glycine chains are from 4 to 15. The test results are shown in Figure.\ref{TIME_SCALE}. The results of computation show that the time scale is $O(N^2)$-$O(N^3)$, about $O(N^{2.7})$ for 3D systems and $O(N^{2.4})$ for 1D systems. The BTD grids are generated in a linear scaling way, and the THC kernel is calculated on a quadratic scale. In systems with a number of basis functions smaller than 700, the main time-consuming step is the building of the THC kernel whose time scale is $O(N_gN_KN_\text{aux})$ without screening. As the size of the system increases, prescreening reduces the time scale to $O(N_KN_\text{aux})$. After the crossing point as 800 basis functions, the computational bottleneck is the calculating the inversion of the overlap fitting matrix $S_{KL}^+$ in Eq.\ref{overlap_S}, which has a cubic scale. To reduce the time consumption of matrix inversion, the overlap matrix can be expanded as a Taylor series. Only the block diagonal parts of $S_{KL}$ are calculated exactly, the contribution of off-diagonal elements is considered in perturbation with second order. Or more efficient hardware such as GPU could be used. This work is currently in progress.

\subsection{Accuracy and efficiency for HF exchange}
For both accuracy and efficiency, JADE is used for parameter optimization. The elements included are C, H, O, N, and F in this work. The molecules in the training set are obtained from G3/99. \cite{G3} The error function is defined as 
\begin{equation}
    F = \sum_{i\in\text{train}}e_i/N + 0.1\times\max_{i\in\text{train}}\{e_i\}
\end{equation}
Here, $e_i$ is the error per atom compared to the Hartree-Fock result calculated by RI-JK for the $i_\text{th}$ molecule in the training set, and N is the number of molecules in the training set. Optimization is carried out with fixed parameters $\{l_1,l_2,l_3,l_4,l_5\}$ as $\{14, 26, 50, 110, 50\}$. The JADE training process is shown in Figure \ref{fig:enter-label}. The minimum of mean absolute deviation (MAE) is 0.0233 kcal/mol per atom. The recommendation for parameters is shown in Table S1.

\begin{figure*}
    \centering
     \subfloat[time scale]{
        \includegraphics[scale=0.455]{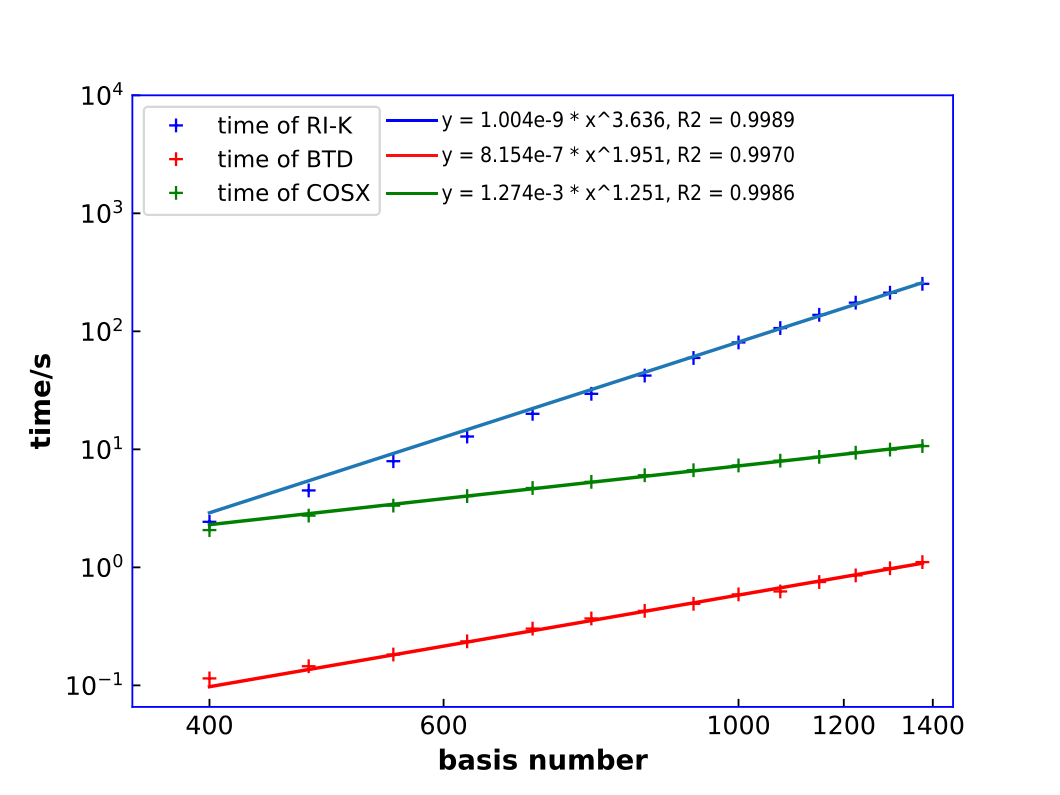}
      }
      \subfloat[time scale]{
        \includegraphics[scale=0.5]{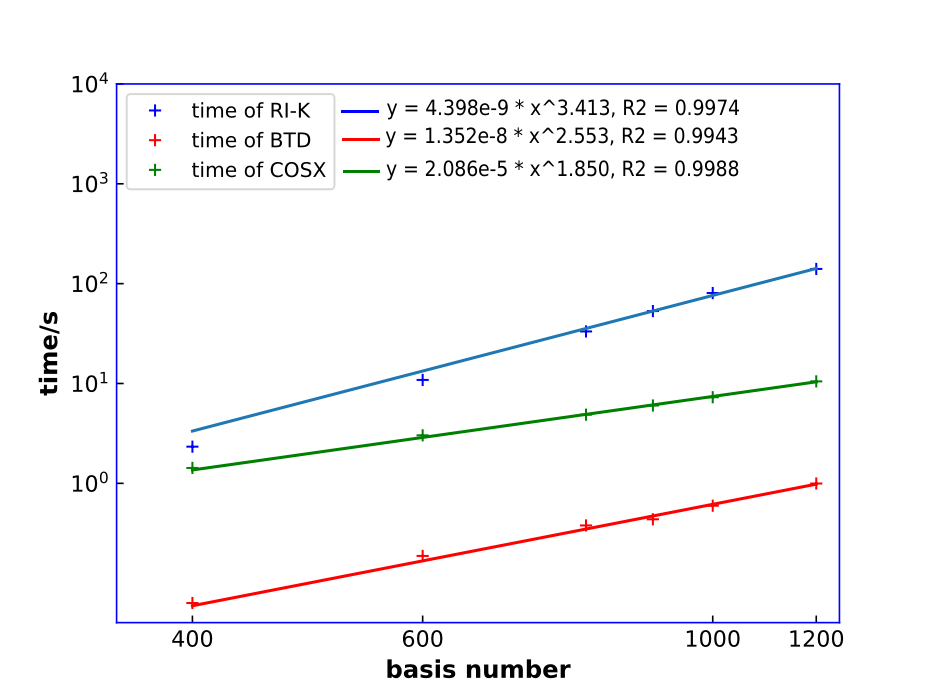}
      }\\
      \subfloat[error]{
        \includegraphics[scale=0.45]{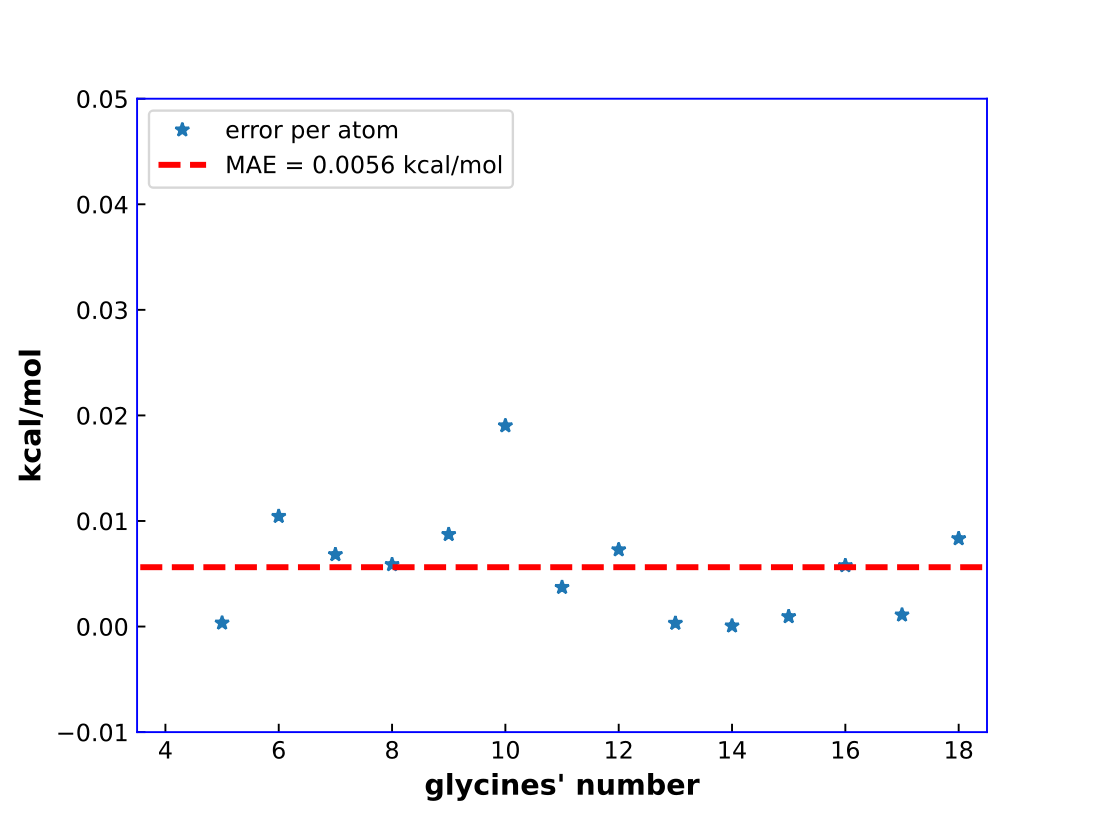}
      }
      \subfloat[error]{
        \includegraphics[scale=0.5]{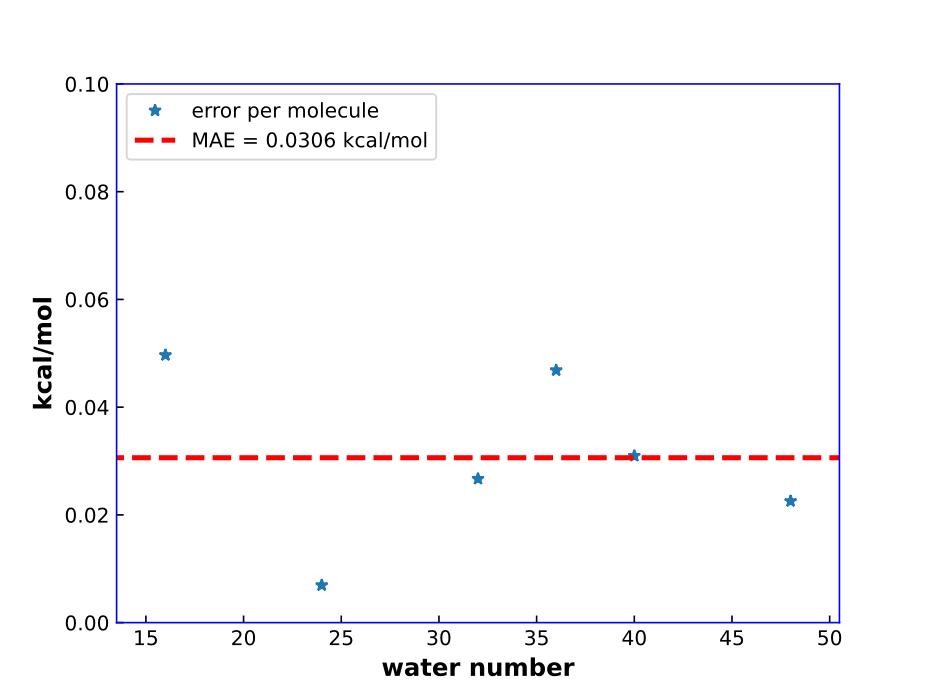}
      }
      \caption{The testing results for glycine chains and water clusters (a,b) time consumption by RI-K, COSX and BTD. (c,d) error per atom of BTD compared to RI-K's results. All calculations use the cc-pVDZ basis set, the cc-pVDZ-JKfir auxiliary basis set for RI-JK and BTD's calculations and cc-pVDZ-RI for RI-COSX's coulomb calculation. }
      \label{exchang_test}
\end{figure*}
Using optimized parameters, testing for exchange calculations is done for glycine chains and water clusters under cc-PVDZ with 32 threads. The number of basis functions, auxiliary functions and interpolative grids are shown in Table S2. The results are shown in Figure \ref{exchang_test} and raw data is shown in Table S3. In \ref{exchang_test}(a), the BTD results are compared to the RI-K and COSX results. Compared to RI-K with the $O(N^4)$ time scale, both COSX and BTD get much better results with 10 times and 100 times speed-up, respectively. In BTD, the calculation of the exchange matrix only takes less than 1 second for the system with 1000 basis functions. The time scale of BTD is about $O(N^{2.0})$ for 1D systems and $O(N^{2.6})$ for 3D systems, while RI-K and COSX are quartic and linear scaling, respectively. In RI-J combined BTD-K calculations for testing systems, the computational bottlenecks are the generation of the THC kernel and the calculation of the Coulomb matrix and the time consumption of exchange calculation in each step can be ignored. Profiting from JADE optimized parameters, the error per atom introduced by BTD is about 0.01 kcal/mol compared to exact RI-K results as shown in Figure \ref{exchang_test}(b). Compared to COSX, BTD performs a higher time scale and may take longer for systems with 10000 basis functions. To make BTD linear scaling, more efficient blocking and screening schemes need to be developed.
\begin{figure*}
    \centering
     \subfloat[time scale]{
        \includegraphics[scale=0.5]{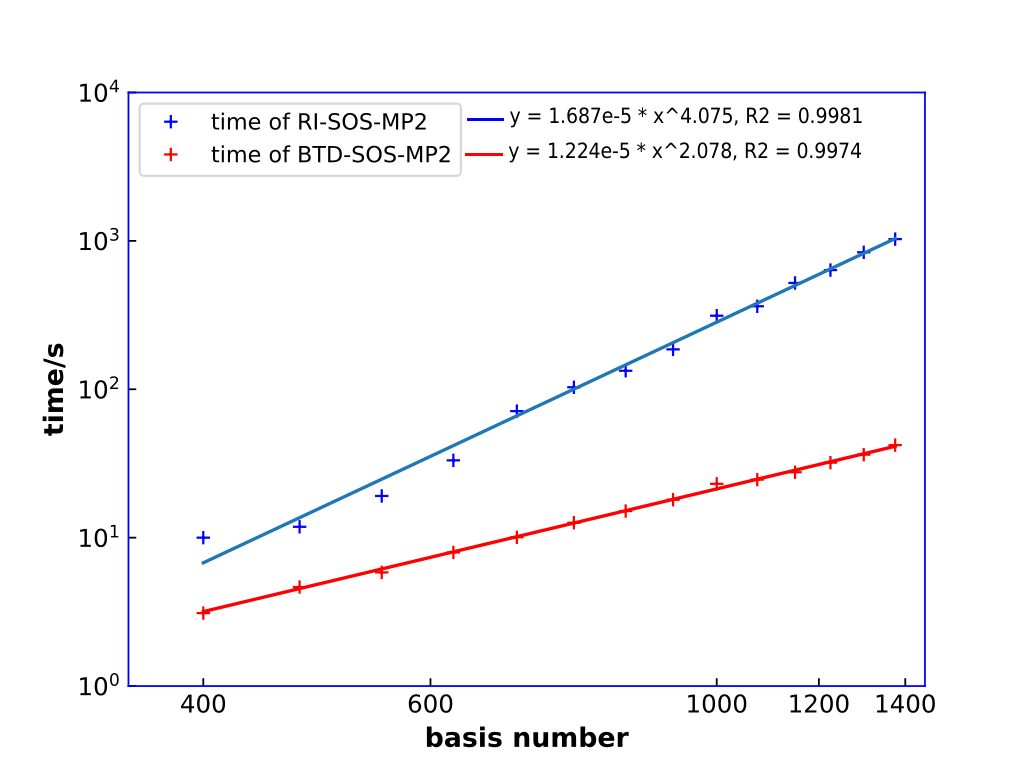}
      }
      \subfloat[time scale]{
        \includegraphics[scale=0.515]{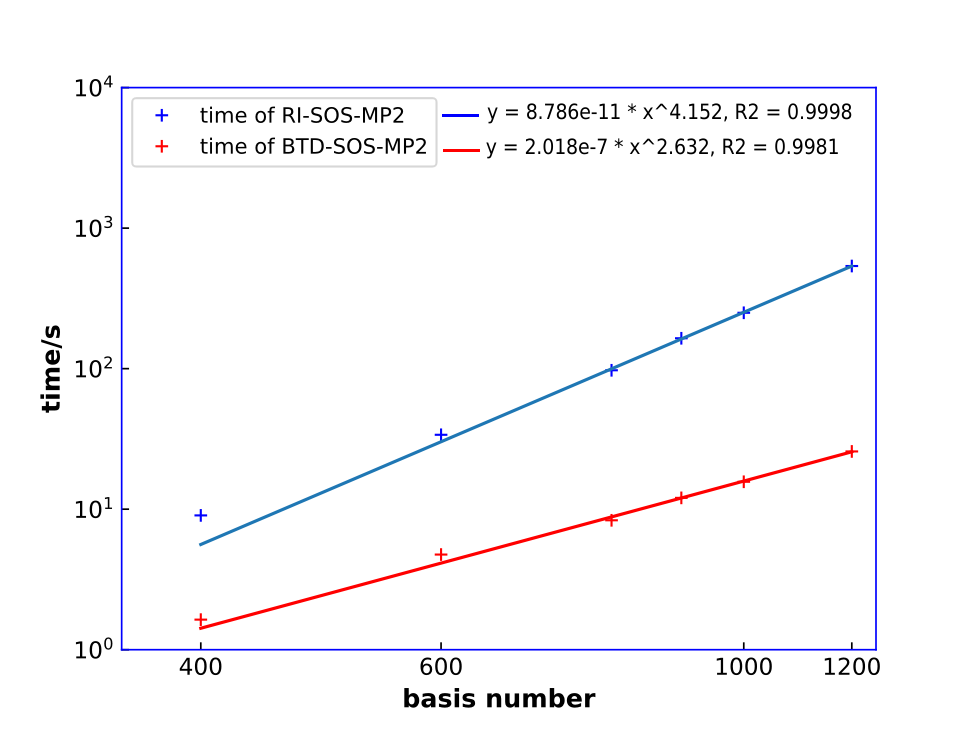}
      }\\
      \subfloat[error]{
        \includegraphics[scale=0.5]{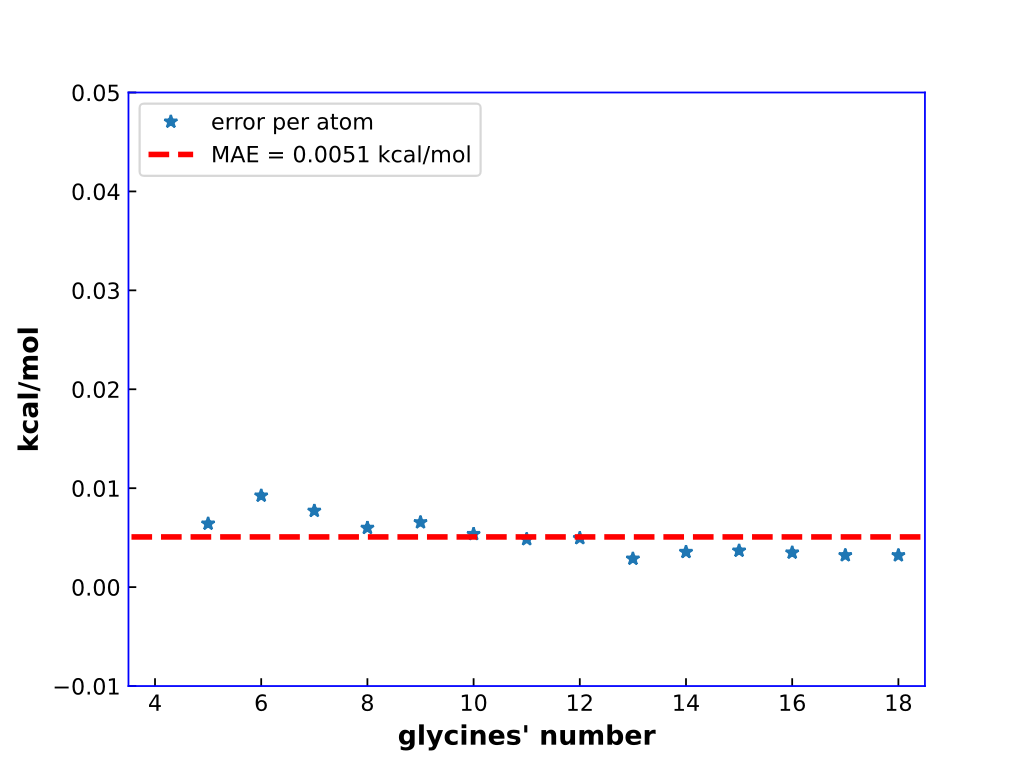}
      }
      \subfloat[error]{
        \includegraphics[scale=0.515]{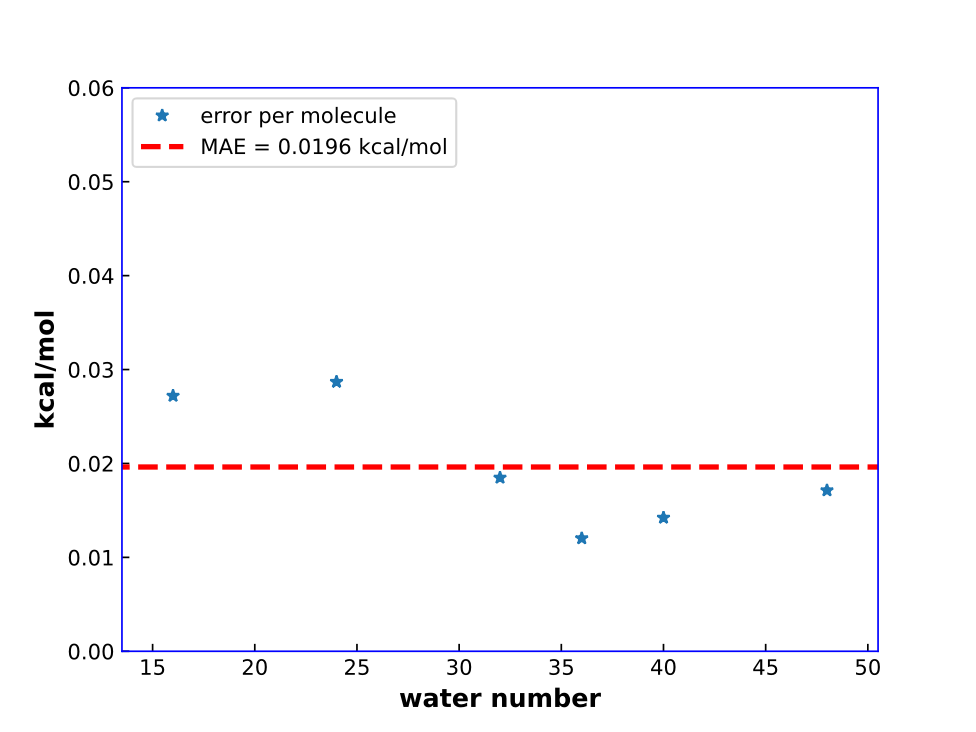}
      }
      \caption{The testing results for glycine chains and water clusters (a,b) time consumption by RI-SOS-MP2 and BTD-SOS-MP2. (c,d) error per atom of BTD compared to RI's results. All calculations use the cc-pVDZ basis set, the cc-pVDZ-RI auxiliary basis set is used. }
      \label{sos-mp2}
\end{figure*}
\subsection{Accuracy and efficiency for SOS-MP2}
SOS-MP2 was also performed for glycine chains and water clusters under cc-PVDZ with 32 threads. The results are shown in Figure \ref{sos-mp2} and raw data is shown in Table S4. For 1D systems, BTD-based SOS-MP2 shows a nearly quadric time scale as $O(N^{2.1})$, while the RI-SOS-MP2 time scale is $O(N^{4.1})$. The error introduced by the BTD approximation is 0.0056 kcal/mol per atom. In the calculation of glycine chains, the most time-consuming step is $O(N_K^2N_\text{aux})$ scaling step as $\sum_{K}B_{KM}P_{KL}(\tau)$. As in the section II D's analysis, this step can achieve the $O(N^2)$ scale by screening based on the local property of Green's function. The time scale of water clusters' results show a similar time scale to Toddo's atomic orbital-based THC-SOS-MP2. \cite{thc-sos-mp21,thc-sos-mp22} The error per molecule is about 0.02 kcal/mol. The parameters used for SOS-MP2 are based on optimized fine-tuning exchange results with fewer grids and higher threshold $\epsilon$ for the Cholesky cutoff. If a specific optimization is performed for SOS-MP2, the deviation would be further reduced.  

\section{Conclusion}
Taking advantage of the dual grid scheme, an algorithm named BTD is achieved with a formal $O(N^3)$ time scale for the low-rank THC-like decomposition of 4c2e-integrals compared to the $O(N^4)$ RI scheme. Using the sparse map of interpolative grids, time consumption of the most time-consuming step can be further reduced to $O(N^2)$. For interpolative grid generation, the pivoted Cholesky decomposition is combined with Hilbert sorting to make a linear-scaling algorithm to make non-redundant grids. The test results show that BTD has a time scale between $O(N^2)$ and $O(N^3)$. The BTD's computational bottleneck suffers from matrix inversion and production. To make the implementation more efficient, hardware efficient algorithm on GPU and matrix inversion scheme based on space sparsity need to be developed.

BTD is combined with Hartree-Fock exchange and SOS-MP2 correlation. To make a good efficiency and maintain accuracy, the parameters in the BTD are optimized by JADE. The time scales of exchange and SOS-MP2 are both $O(N^2)$ to $O(N^3)$. Prescreening is used to reduce computations of block pair without distribution. The test results show that the BTD based algorithm is efficient for both 1D and 3D systems. To calculate exchange in linear scaling way, a more efficient blocking scheme and prescreening should be applied.

We notice that BTD calculation only needs 1c2e-integrals and the basis function's information is included in the auxiliary function and real-space grids. These properties prevent BTD from calculation of complex two-electron integrals and are insensitive to the format of the basis function. BTD is easy to be extended to GPU algorithm and more general local basis function as Slater orbitals. RPA and $GW$ can also be combined with BTD to achieve a formal $O(N^3)$ scheme, relative works are in progress.
\begin{acknowledgments}
    This project is supported by the National Natural Science Foundation of China (Nos. 22473092, 22173076, 22373077).
    \end{acknowledgments}
    
    \section*{AUTHOR DECLARATIONS}
    \subsection*{Conflict of Interest}
    The authors have no conflicts to disclose.
    \section*{Data Availability Statement}
    See supplementary material associated with this article, including Tables S1-S4.
    
    \section*{reference}
    \bibliography{aipsamp}
\end{document}